\documentclass[epsfig,12pt]{article}
\usepackage{amsmath,epsfig,amssymb,latexsym,cite,xy,color}
\xyoption{matrix}\xyoption{arrow}\xyoption{frame}
\usepackage{shadow}

\newcommand{\newsection}[1]{
\addtocounter{section}{1} 
\setcounter{subsection}{0} \addcontentsline{toc}{section}{\protect
\numberline{\arabic{section}}{{\rm #1}}} \vglue .0cm \pagebreak[3]
\noindent{\large \bf  \thesection. #1}\nopagebreak[4]\par\vskip .3cm}
\newcommand{\newsubsection}[1]{
\addtocounter{subsection}{1}
\addcontentsline{toc}{subsection}{\protect
\numberline{\arabic{section}.\arabic{subsection}}{ #1}} \vglue .4cm
\pagebreak[3] \noindent{\sc \thesubsection.
#1}\nopagebreak[4]\par\vskip .3cm}

\newcommand{\is}{\! &\! =\! & \!}
\newcommand{\nonu}{\nonumber\\[1mm]}
\newcommand{\GG}{\Gamma}

\newcommand{\Ext}{{\rm Ext}}

\newcommand{\bu}{\overline u}
\newcommand{\bd}{\overline d}
\newcommand{\bbe}{\overline e}

\newcommand{\bnu}{\overline \nu}
\begin{document}

\addtolength{\baselineskip}{.68mm}
\newlength{\extraspace}
\setlength{\extraspace}{1.2mm}
\newlength{\extraspaces}
\setlength{\extraspaces}{1.2mm}
\newcounter{dummy}

\newcommand{\be}{\begin{equation}}
\addtolength{\abovedisplayskip}{\extraspaces}
\addtolength{\belowdisplayskip}{\extraspaces}
\addtolength{\abovedisplayshortskip}{\extraspace}
\addtolength{\belowdisplayshortskip}{\extraspace}
\newcommand{\ee}{\end{equation}}

\newcommand{\ba}{\begin{eqnarray}}
\addtolength{\abovedisplayskip}{\extraspaces}
\addtolength{\belowdisplayskip}{\extraspaces}
\addtolength{\abovedisplayshortskip}{\extraspace}
\addtolength{\belowdisplayshortskip}{\extraspace}
\newcommand{\ea}{\end{eqnarray}}

\begin{titlepage}
\begin{center}

{\hbox to\hsize{
\hfill PUPT-2165}}

{\hbox to\hsize{\hfill hep-th/0508089}}

\vspace{3.5cm}

{\Large \bf Building the Standard Model on a D3-brane}\\[1.5cm]

{\large Herman Verlinde\footnote{{\tt verlinde@princeton.edu}} and Martijn Wijnholt\footnote{{\tt wijnholt@princeton.edu}} }\\[8mm]

{\it Department of Physics,\\[2mm] Princeton University,\\[2mm] Princeton,
NJ 08544}\\[6mm]

\vspace*{3.5cm}

{\bf Abstract}\\

\end{center}
\noindent
We motivate and apply a bottom-up approach to string phenomenology,
which aims to construct the Standard Model as a decoupled world-volume theory
on a D3-brane. As a concrete proposal for such a construction, we consider
a single probe D3-brane on a partial resolution of a del Pezzo 8 singularity.
The resulting world-volume theory reproduces the field content and interactions 
of the MSSM, however with a somewhat extended Higgs sector. An attractive 
feature of our approach is that the gauge and Yukawa couplings are dual to
non-dynamical closed string modes, and are therefore tunable parameters.

\end{titlepage}

\newpage

\tableofcontents

\newpage

\definecolor{gray}{rgb}{.5,.5,.5}

\def\IC{{\bf C}}
\def\IZ{{\bf Z}}
\def\IP{{\bf P}}
\def\lR{{\bf R}}
\def\lZ{{\bf Z}}
\def\lP{{\bf P}}

\vspace{-9mm}

\newsection{ Introduction: Bottom-Up String Phenomenology}

String theory has presented itself with the formidable task of taming the
quantum realm of gravity, while simultaneously furnishing a  predictive and
testable theory of particle physics. In attempting to meet  this dual challenge,
string phenomenology traditionally adopts a ``top-down'' point of view,
which aims to construct realistic compactification scenarios starting from the
full 10-dimensional closed string theory \cite{gsw}\cite{burtea}, 
possible augmented with one or more
D-branes \cite{D}. The thrust of this approach is that, by simultaneously
controlling and scanning both the string scale geometry and the low energy field theory,
one can isolate realistic backgrounds that meet all consistency requirements
at both ends.  Recent progress in mapping out the ``closed string theory landscape,''
the vast collection of potentially stabilized string vacua \cite{KKLT}, has
strengthened the belief that such consistent backgrounds indeed exist, possibly
even in abundance \cite{toomany}. Finding a single one of them, however, still seems
far too challenging a task at present.

In a complementary development, fueled by the deepened understanding of string
dualities and D-brane physics, open string theory has evolved into a remarkably
successful tool for building 4-d supersymmetric field theories.  In particular,
it is now realized that by taking a judicious low energy limit of the world-volume
theory on $N$ D3-branes, one recovers a purely 3+1-dimensional
gauge theory, decoupled from gravity and higher dimensional dynamics \cite{Malda}.
In this decoupling limit, the closed string
background gets frozen into a set of non-dynamical, and thus largely tunable,
gauge invariant couplings. By placing
one or more D-branes near various types of geometric singularities, realizations of
 large classes of gauge theories have been uncovered
 \cite{ALE,KS,albion,MoPl,KW,Greene,ahe,Muto}.
Evidently, open string theory has become the preferred duality frame for
representing weakly coupled, as well as strongly  coupled, 4-d quantum field theories.

Given this rich ``open string theory landscape,'' it is a well-motivated question whether,
with currently available technology, one can find an explicit realization of the supersymmetric
Standard Model as the world-volume theory on one or more
D3-branes. Since every decoupled theory, via its space of tunable couplings,
stretches out over a sizable open neighborhood within the space of 4-d field
theories, one can even aim to reproduce the spectrum and couplings
within phenomenological bounds.
Though clearly a non-trivial  challenge, this question is still far less ambitious,
and thus easier to answer, than finding a fully realistic closed string compactification.
But it would be a useful first step: only after one knows how to
represent the observed particle spectrum as an open string theory near a suitable singularity,
one can start to look for compact geometries that contain this singularity. We thus view the
bottom-up approach to string phenomenology \cite{hv,bottomup,Frampton,urangareview,SMb,bcls,bjl,Grana,Cascales}
as a promising route towards unlocking some of the mysteries
of the closed string theory landscape.

The experimental fact  that guides the bottom-up perspective is the exponential separation
between the TeV scale of particle physics and the Planck scale of quantum gravity.
Warped compactifications 
\cite{hv}\cite{GKP} provide a geometric implementation of the gauge hierarchy, via the
assumption that all low energy physics takes place in a highly red-shifted
region of the internal geometry.
This geometric viewpoint thus naturally places the Standard Model on a world-brane near the
apex of a warped throat.

A key manifestation of the gauge hierarchy is that, at the TeV scale, one can make a clean
separation between dynamical gauge and matter fields and non-dynamical coupling
constants -- even though both start out as equally dynamical degrees of freedom in the full high energy
string theory. It is thus accurate, and even appropriate,
to isolate the low energy worldbrane physics from the closed string dynamics, by taking
a decoupling limit in which the Planck scale is sent off to infinity.
In geometric terms, this limit replaces the finite warped throat region by an infinite,
non-compact Calabi-Yau  singularity.

\begin{figure}[htbp]
\begin{center}
\leavevmode\hbox{\epsfxsize=14cm \epsffile{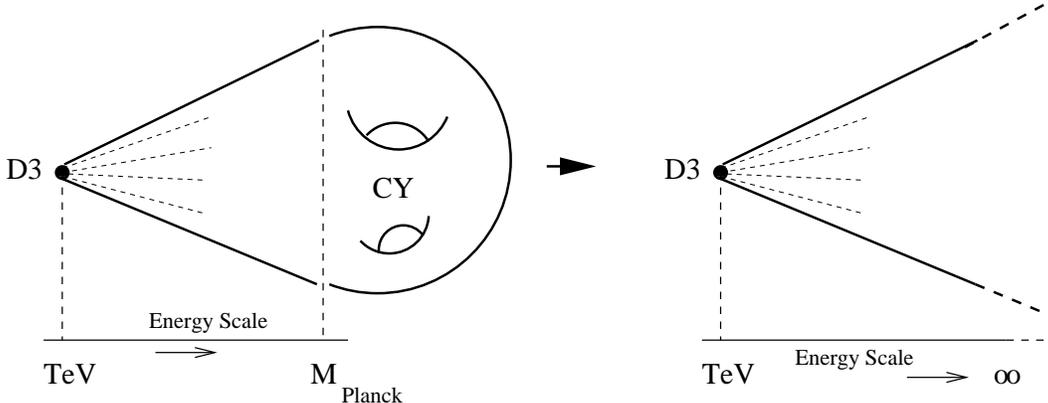}}\\[3mm]
\caption{\textcolor{black}{Our bottom-up approach to
string phenomenology assumes that the Standard Model
is localized on a D3-brane at the apex of a highly warped throat. The
D3-brane theory is accurately described via a decoupling limit, in which the
Planck scale is sent off to infinity, leaving behind a non-compact Calabi-Yau singularity. }}
\end{center}
\end{figure}
Due to the interaction with the ambient geometry, a D3-brane on a CY singularity breaks up
into various fractional branes. As a result, its world-volume theory takes the non-trivial
form of a quiver gauge theory:
it has one $U(n_i)$ gauge multiplet for each constituent
fractional brane and bifundamental chiral matter
associated with each brane 
intersection  \cite{ALE,KW,resol, bobby,mike,mikefiol,dido,mayr,Tomasiello,
fhh1,freddy,Katz,asp,martijn,chris,seimut}.
In the decoupling limit,  the gauge invariant coupling constants of this quiver gauge theory
are determined by non-dynamical asymptotic boundary conditions on the closed string fields,
and can thus be viewed as continuously tunable parameters.

This paper is our first progress report on our search for a
D3-brane realization of the Standard Model. Our strategy is as
follows. We start  by setting up the rules that define quiver
gauge theories, and introduce a corresponding minimal quiver
extension of the MSSM, which we call the MQSSM. Our target is to
find its geometric dual. Since it is hard to immediately guess the
right geometry, we first identify a class of Calabi-Yau singularities,
such that the probe D3-brane theory is just large enough to
contain the MSSM, and has a rich enough space of couplings and
vacua to allow the necessary tuning. We then look for a suitable
symmetry breaking process towards the MQSSM quiver theory. After
translating into the dual geometric language, the symmetry
breaking amounts to into a specific partial resolution of the CY
singularity, which then provides the sought after
geometric dual. To go further one must turn on various soft supersymmetry
breaking terms. Apart from some general comments, we leave this 
problem for the future.


The specific class of geometries we will consider are the del Pezzo 8 singularities.
By following the outlined procedure, we identify a specific partial resolution of the
del Pezzo 8 geometry for which the D3-brane gauge theory has the Standard Model
gauge group, $SU(3)_C \times SU(2)_L \times U(1)_Y$,
and matter content, three families of quarks and leptons with all the right charges,
 plus a somewhat extended Higgs sector.
All matter fields appear with the proper chiralities and all have
classically tunable Yukawa couplings. The quiver diagram of the
model is given in fig 8. In the final section, we discuss some
physical aspects of the MQSSM and address some possible criticisms of 
our approach.

\bigskip
\bigskip

\newsection{  A Quiver Extension of the MSSM }

It will be useful to
introduce a minimal quiver gauge theory extension of the supersymmetric
Standard Model.  The motivation for presenting it is two-fold: (i) it will help with
recognizing, among the vast collection of possibilities, those open string theory
constructions that may contain the MSSM as a special limit, and (ii) it will
give a useful preview of typical extra features that arise in generic
open string set-ups, and that need to be dealt with in making a fully
realistic model. For both reasons,  let us adopt the quiver diagrammatic rules
that apply to D3-branes on CY singularities. These are (see next section):

(a) Each node of the quiver represents a gauge multiplet with $U(k)$ gauge symmetry.

(b) Each oriented line between two nodes represents a
bi-fundamental chiral multiplet.

(c) There is an equal number incoming and outgoing lines connected to every node.

\noindent
As we will see, these characteristics all have a direct
geometric origin. Rule (c) in particular ensures
the absence of non-Abelian gauge anomalies for the case of multiple D3-brane probes.
Two additional rules, that apply to an  especially convenient class of D3-brane configurations,
known as ``exceptional collections,''  are:

(d) There are no lines that begin and end at the same node.

(e) There is only one type of oriented lines between any pair of nodes.

\noindent
Rule (d) excludes the presence of adjoint matter multiplets. Rule (e) states that all
bifundamental matter multiplets are purely chiral.

\medskip

\begin{figure}[htbp]
\begin{center}
\leavevmode\hbox{\epsfxsize=9cm \epsffile{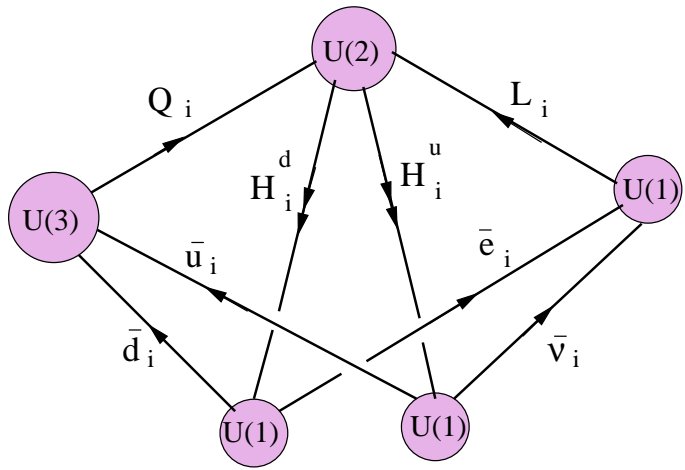}}\\[3mm]
\caption{{The MQSSM is the minimal quiver extension of the Standard Model, as obtained
via the rules (a)-(e).}}
\end{center}
\end{figure}

The minimal quiver extension of the MSSM, drawn by using the five rules (a) through (e),
is given in figure 2.  It depicts all the gauge charges of the fields, while each closed triangle
of the diagram represents a possible Yukawa coupling.
We see that relative to the MSSM, there are several extra $U(1)$ factors,
and a number of extra Higgs doublets: two pairs for each
generation. The additional Higgses are forced on us by rule (c) and the requirement of having all
the expected supersymmetric Yukawa couplings

Let us briefly discuss the $U(1)$ factors. We call the node on the right $U(1)_0$ and
the two nodes at the bottom $U(1)_u$ and $U(1)_d$. The five $U(1)$ generators
are denoted by $\{Y_0, Y_1^u,Y_1^d, Y_2, Y_3\}$.
The charges of the matter fields are given in the table below.

We note that the total sum $Q_{tot} = \sum_i Y_i$ automatically 
decouples: none of the fields is charged under $Q_{tot}$.  Of the remaining four
generators, some are anomalous. The anomalies cancel for the combinations
\be
B-L= {1\over 3} Y_3-Y_0  \qquad \qquad Y = {1\over2} (Y_1^d - Y_1^u -Y_0) + {1\over 6} Y_3
 \ee
Obviously,  only $U(1)_Y$ represents an actual local symmetry in the MSSM. The
$U(1)_{B-L}$ symmetry has the desirable
consequence that it helps suppress the rate of proton decay, but obviously, it needs to be
spontaneously broken at an energy scale of order of a TeV or larger.
A natural symmetry breaking mechanism
would be to assume a non-zero vacuum expectation value for the 
bosonic superpartners of the right-handed neutrino fields, which would also mesh well
with the small neutrino Yukawa couplings.

\medskip

{{\small \begin{equation}
\begin{array}{|c||r|r|r|r|r|r| |c| }
\hline
& \ Y_0\  & \ Y_1^d \  & \ Y_1^u\  & \ Y_2\ & \ Y_3\  & \ \ Y \\[1mm] \hline \hline
Q & 0 & 0 & 0 & -1 & 1  & {1\over 6} \\[1mm] \hline
\bu & 0 & 0 & 1 & 0 & -1 & - {2\over 3} \\[1mm] \hline
\bd & 0 & 1 & 0 & 0 & -1 & {1\over 3} \\[1mm] \hline
L & 1 & 0 & 0 & -1 & 0 & - {1\over 2}  \\[1mm] \hline
\bnu & - 1 & 0 & 1 & 0 & 0 & 0   \\[1mm] \hline
\bbe & - 1 & 1 & 0 & 0 & 0 &  1 \\[1mm] \hline
H^{u} & 0 & 0 & -1 & 1 & 0 & {1\over 2}  \\[1mm] \hline
H^{d}  & 0 & -1 & 0 & 1 & 0 & - {1\over 2} \\[1mm] \hline
\end{array}
\label{eq:charges}
\end{equation}}
\vspace{0mm}
{{\begin{center}{Table 1. The $U(1)$ charges}\end{center}}}
\vspace{0mm}

\medskip

The remaining two $U(1)$'s indeed have mixed anomalies. In string theory realizations of
quiver theories, these are canceled via a generalized Green-Schwarz
mechanism.  Moreover, via the coupling to the RR-forms, the corresponding gauge bosons
typically acquire a mass of order the string scale. We will give a brief outline of this
mechanism in the last section.  From the low energy perspective, these $U(1)$'s thus
survive as anomalous global symmetries, that, among other things, forbid the
presence of $\mu$-terms in the classical superpotential.\footnote{These will have to arise from 
some other source, such as the Giudice-Masiero mechanism \cite{GM}.}

In general, the presence of extra $U(1)$ factors as well as extra Higgs fields is characteristic
of many string theoretic models. Both are acceptable extensions of the Standard Model,
provided the masses and couplings are tuned to satisfy the appropriate phenomenological bounds.
For now, however, we postpone the discussion of these issues:
instead we set out to find a Calabi-Yau singularity such that the
world-volume theory of a probe D3-brane reproduces the MQSSM, the quiver gauge theory of fig 2.

\bigskip
\bigskip

\bigskip

\newsection{ D3-brane on a del Pezzo 8 Singularity}

In this section we introduce the quiver gauge theory on the world volume of a
D3-brane on a del Pezzo 8 singularity. This quiver theory has been previously derived in \cite{martijn}
using a geometric description,
and in this section
we begin by recalling some
elements of that  construction.\footnote{More details and additional work on D-branes at Calabi-Yau
singularities can be found the original 
literature \cite{mike,mikefiol,dido,mayr,Tomasiello,fhh1,freddy,asp,Katz,martijn,chris,seimut}.}
The geometric description is strictly only accurate for the holomorphic
F-term data of the quiver theory.
It relies on the fact that the (complexified) K\"ahler moduli only appear in the D-terms, so that
we can  extrapolate to large volume without affecting the F-terms. In the large volume limit,
topological data
of the quiver gauge theory, such as the gauge group, the number and
representations of the matter multiplets, as well as the holomorphic superpotential,
can be accurately obtained via the geometrical methods outlined below.

\bigskip

\newsubsection{Geometry of del Pezzo 8}

A del Pezzo surface is a manifold of complex
dimension 2, with a positive first Chern class. It is labelled by an integer
$n$; the $n$-th del Pezzo surface ${\cal B}_n$ can be  represented as either
$\IP^2$ blown up at $n \leq 8$ generic points, or as $\IP^1 \times \IP^1$ blown
up at $n-1$ points. We choose the first representation.
Here blowing up a point means replacing it by a sphere.
By placing the canonical line bundle over ${\cal B}_n$,
one obtains a non-compact Calabi-Yau three-fold. In the limit where the Del Pezzo surface
shrinks to zero size, one obtains a singular three-fold. We will call this the del Pezzo $n$ singularity.

The second Betti number $b_2({\cal B}_n)$ is equal to $n+1$.
A basis of $H_2({\cal B}_n,{\lZ})$ is given by the
hyperplane class $H$ in ${\lP}^2$ %
plus one generator $E_i$ for each of the $n$ blown up points.  The generators $E_i$ are
called exceptional curves.  The intersection numbers are given by
\be
\label{intp}
H\cdot H \, = \, 1, \qquad  E_i\cdot E_j =  - \delta_{ij} , \qquad H \cdot E_i = 0\, .
\ee
The canonical class of the del Pezzo surface is
\be
\label{canon}
K = -3 H + \sum_{i=1}^n E_i
\ee
It has self intersection $K\cdot K  = 9 - n$.  The first Chern class of ${\cal B}_n$ is $c_1(T{\cal B}_n)=-K$.
The characteristic property of a del Pezzo surface is that $c_1$ is ample, that is, it has
positive intersection with every effective curve on ${\cal B}_n$. This in particular implies
that $K$ must have positive self-intersection, which gives the restriction $n \leq8$.

In this paper, we will mostly consider the 8-th del Pezzo surface ${\cal B}_8$. It
can be constructed as a hypersurface of degree six in the weighted projective space
${\bf WP}^3_{1,1,2,3}$ with homogeneous coordinates $(x,y,z,w)$, defined via an equation
of the generic form
\be
\label{dpeight}
w^2 = A z^3 + B y^6 + C x^6 + \ldots
\ee
This defines a del Pezzo surface because the sum of the weights exceeds the degree of
the surface.
The 2-d homology of ${\cal B}_8$ is generated by 8 exceptional curves $E_i$, corresponding to
the 8 blow up points, and the hyperplane class $H$. Via the intersection pairing (\ref{intp}), 
$H_2({\cal B}_8,{\bf Z})$ takes the form  of an integral lattice in ${\bf R}^9$. 
It has the remarkable property that
the 8-dimensional degree zero sub-lattice, defined as those elements with zero intersection with
$c_1=-K$,  is even and unimodular. In other words, it is isomorphic to the root lattice of $E_8$.
The $8$ simple roots, all with self-intersection -2, can be chosen as follows
\be
\label{ddd}
\alpha_i = E_i - E_{i+1},   \ \ \ \ \ \mbox{\footnotesize $i=1,\ldots, 7$}
\qquad \qquad \alpha_8 = H-E_1-E_2-E_3
\ee
Combined with $K$, they form a complete basis of $H_2({\cal B}_8,{\bf Z})$, with intersection form
\be
\alpha_a \cdot \alpha_b = - A_{ab} \, ,
\qquad \qquad K \cdot \alpha_a = 0
\ee
with $A_{ab}$ the Cartan matrix of $E_8$.  The 2-cycles $\alpha_i$ can thus be
identified with nodes on the $E_8$ Dynkin diagram, as drawn below. 
This identification between 2-cycles and simple roots gives rise to a natural action of the
Weyl group of $E_8$ on $H_2({\cal B}_8,{\bf Z})$, in terms of global diffeomorphisms
on ${\cal B}_8$ that exchange the exceptional curves while preserving $K$.
The Weyl reflections in the simple roots $\alpha_i$ with $i\leq 7$ are simply diffeomorphisms
that interchange two of the blown up points, while keeping the rest of the surface fixed. The
reflection in $\alpha_8$ looks more complicated, but can be understood in a similar manner.

\begin{figure}[htbp]
\begin{center}
\leavevmode\hbox{\epsfxsize=5.5cm \epsffile{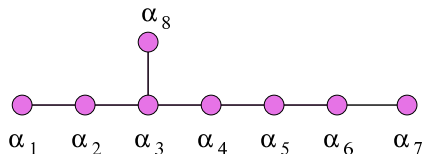}}\\[3mm]
\caption{ \textcolor{black}{The 2-cycles $\alpha_i$ are identified with nodes on
the $E_8$ Dynkin diagram.}}
\end{center}
\end{figure}

The ellipses in the homogeneous equation (\ref{dpeight})
represent additional terms that deform the complex structure of the del Pezzo 8 surface.
An elegant description of this space  of complex structure deformations has been given in \cite{morgan}.
It is built on the observation that  the homogeneous eqn (\ref{dpeight}) can chosen of the form
\vspace{2mm}
\be
\label{mfw2}
w^2 = 4z^3 - g_2 \, z y^4 - g_3 \, y^6 \, + \, P(x,y,z),
\ee
where $P(x,y,z)$ is a suitable homogeneous polynomial \cite{morgan} that vanishes at $x\! =\! 0$.
The degree one surface $x\! =\! 0$ is an anti-canonical divisor, and defines an elliptic curve ${\cal E}$, 
given by a Weierstrass equation 
in ${\bf WP}_{1,2,3}^2$. The space of continuous
complex structure deformations that
keep ${\cal E}$ fixed is 8 dimensional.\footnote{A direct way to interpret these 8 complex
structure deformations is that they parameterize the locations of 4 of the 8 points
that are blown-up to produce ${\cal B}_8$ from ${\bf P}^2$. The positions of the four other points
do not give rise to complex structure moduli, since they can be held fixed by using the
$PGL(3,{\bf C})$ group of coordinate transformations of the underlying  ${\bf P}^2$.}
Natural coordinates on this space are the parameters
that specify the polynomial $P(x,y,z)$.\footnote{As it turns out \cite{morgan}, these
describe homogeneous coordinates on the weighted projective space:
${\bf WP}^8_{1,2,2,3,3,4,4,5,6}$. Remarkably,
the set of weights of the projective space coincides with 1 plus the set of Dynkin labels of the
highest coroot of the $E_8$ Kac-Moody algebra.
As explained in \cite{morgan}, the above embedding
of ${\cal E}$ inside the del Pezzo 8 surface 
induces an $E_8$-bundle over ${\cal E}$. The construction
can thus be used to establish an isomorphism between the space of complex structure deformations
of ${\cal B}_8$ and the moduli space of $E_8$ bundles over ${\cal E}$.}

Later on, in our construction of a Standard Model-like gauge theory, we will consider a special
degenerate limit of the del Pezzo 8 geometry, in which some of the 2-cycles $\alpha_i$, given
in eqn (\ref{ddd}), become effective curves on the del Pezzo surface. The del Pezzo surface
then develops a singularity of the
appropriate A-D-E type. The maximally degenerate surface of this type is obtained by
setting $P(x,y,z)=0$ in (\ref{mfw2}). The resulting surface is an elliptic singularity of type $E_8$.
More generally, one can get an $H$-type singularity for
every subgroup $H$ of $E_8$.

\bigskip

\newsubsection{Quiver gauge theory of a D3-brane on del Pezzo 8}

\smallskip

\newcommand{\FF}{F}

Let ${\cal X}$ denote a non-compact Calabi-Yau manifold given by a complex cone
over a collapsing del Pezzo 4-cycle ${\cal B}$.
The D3-brane configurations that we will consider are brane-worlds
that fill the 3+1 flat directions, and therefore are localized at a point in ${\cal X}$. In the strongly
curved background at the tip of the cone, the D3-brane will typically split
into several so-called fractional branes that wrap vanishing cycles in ${\cal X}$.

As far as the F-terms is concerned, we may blow up the vanishing cycles and
perform computations in the large volume limit.
From a large volume perspective,
the geometric characterization of a fractional brane is as a sheaf $\FF_i$, which one can think of
as a bundle supported on the collapsing del Pezzo surface.
The RR-charges of a sheaf $\FF_i$ are combined in the charge vector
\be
{\rm ch}(\FF_i)=({\rm rk}(\FF_i), c_1(\FF_i),{\rm ch}_2(\FF_i)),
\ee
which specifies the (D7,D5,D3) charge of $\FF_i$.
The D7 charge is called the rank ${\rm rk}(\FF_i)$ of the sheaf, while
the D5 charge is equal to the first Chern class $c_1(\FF_i)$ and specifies a
two-cycle around which the D5-component of the fractional brane is wrapped.
If we think of a fractional brane $F_i$ with non-zero rank as a 7-dimensional
gauge theory on a D7-brane,  $c_1(F_i)$ indicates the presence of
non-trivial magnetic flux supported on the corresponding 2-cycles, and
${\rm ch}_2(F_i)$ represents a non-trivial instanton number.

\medskip

In this language, the D3-brane itself is naturally represented as a
sky-scraper sheaf ${\cal O}_p$ localized at a single point $p$. It splits up in a collection
of fractional branes $\FF_i$, each with integer multiplicities $n_i$, such that the
charge vectors of all fractional branes add up to that of a single D3-brane
\be
\label{nsum}
\sum_i n_i \, {\rm ch}(\FF_i) = {\rm ch}({\cal O}_p) = (\, 0\, , \, 0\,, \, 1 \, )
\ee
To satisfy this condition, some of the charges have to be negative, since all charges
associated with 4- and 2-cycles would have to add up to zero. If $n_i$ is negative,
it doesn't necessarily mean that it counts anti-branes. In the limit when
the del Pezzo surface collapses, the central
charge vectors will line up after taking the small volume limit, so that all fractional
branes preserve the same four supersymmetry charges.

\begin{figure}[t]
\begin{center}
\leavevmode\hbox{\epsfxsize=6cm \epsffile{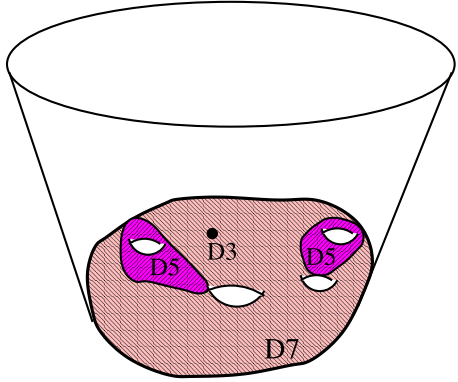}}\\[3mm]
\caption{ {D7, D5 and D3-branes wrapped 4-, 2- and 0-cycles of the internal manifold. }}
\end{center}
\end{figure}

Each type of fractional brane, with
multiplicity $n_i$, contributes a $U(|n_i|)$ factor to the total gauge group of the
world-volume theory.  The corresponding ${\cal N}=1$ gauge multiplet is furnished by
the lightest modes of open strings with end-points on the same type of fractional brane.
The massless spectrum of open strings that stretch between two different types of
fractional branes $\FF_i$ and $\FF_j$ represent chiral multiplets that transform
in the bi-fundamental representation of the corresponding $U(|n_i|) \times U(|n_j|)$
gauge group. In case the branes are space filling, i.e. have support
on the whole Calabi-Yau, the massless modes correspond to elements of the cohomology
of the Dolbault 
operator 
acting on the space of bi-fundamental valued anti-holomorphic forms,
$\Omega^{(0,\cdot)}(\FF^*_i,\FF_j)$.  The number of bi-fundamental fields is therefore
counted by the proper generalization to sheaves of the cohomology group
$H^{(0,\cdot)}(\FF_i^* \otimes \FF_j)$, known as the Ext groups
${\rm Ext}^k(\FF_i,\FF_j)$. 
Since our fractional branes are not space-filling, we instead need to
distinguish between a sheaf $\FF$, living on the del Pezzo 4-cycle ${\cal B}$,
and the associated push-forward $i_* \FF$ on the Calabi-Yau ${\cal X}$,
which can be thought of as $F$ extended by zero on ${\cal X}$.
Taking this into account, one concludes\footnote{In short, the argument that leads to this
conclusion is as follows.
Let us denote the normal bundle of the collapsing cycle by $N$.
Then the spectrum of massless modes is counted by
\cite{Witten:1992fb} \cite{Katz}
\be
\Ext^r_{\cal X}(i_* \FF_j, i_* \FF_k) = \sum_{p+q=r} \Ext^p_{\cal B}(\FF_j,\FF_k \otimes \Lambda^q N).
\ee
For a del Pezzo surface, the normal bundle is equal to the canonical line
bundle, $N = K$. Given a generator for $\Ext^p_{\cal B}(\FF_j,\FF_k)$, we can use Serre duality on
${\cal B}$ to get a generator
in $\Ext^{2-p}_{\cal B}(\FF_k, \FF_j \otimes K)$, hence we get {\it two} $\Ext$
generators on the Calabi-Yau ${\cal X}$.
These two generators are in turn related by Serre duality on
${\cal X}$, which maps
 $\Ext_{\cal X}^p(i_* \FF_j, i_* \FF_k)$ isomorphically to $\Ext_{\cal X}^{3-p}(i_* \FF_k, i_* \FF_j)$.
There is a simple physical interpretation for this doubling. The degree (mod 2) of the $\Ext$ group is
related to 4-d chirality
through the GSO projection. Two generators related by Serre duality therefore have opposite chirality
and opposite bifundamental  charge,
and so they give rise to a particle
and its corresponding antiparticle. Since by convention chiral superfields contain a left-handed
spinor, the dual pair of generators gives  a single chiral superfield in four dimensions --
the second generator descends to the conjugate anti-chiral superfield.} that for
each generator of $\Ext^p_{\cal B}(\FF_j,\FF_k)$, one has 
exactly one chiral field in four dimensions. This
is all we need to know for now. 

On a given  Calabi-Yau singularity, there are different possible choices of basis for the
fractional branes. The type of basis that is most well-understood are the so-called
``exceptional collections''.  These satisfy the special criteria that

(i) ${\rm Ext}^m(\FF_i,\FF_i) = 0$ for $m>0$. This implies the absence of adjoint matter
for a collapsing del Pezzo.

(ii) there exist an ordering of the $F_i$'s, such that ${\rm Ext}^m(\FF_i,\FF_j)=0$ for all but one $m$ if
${}$~~~~~~~~~~$j>i$ and for all $m$ if $i>j$.

\noindent
The second condition (ii) implies that  the bi-fundamental
multiplets between any two given nodes has only one type of chirality.

\smallskip

Let us specialize to the case of the del Pezzo 8 singularity. The
total homology of ${\cal B}_8$ is 11-dimensional; we thus expect
to find 11 types of fractional branes.  Mathematicians have
identified a natural choice of basis of coherent sheaves on del
Pezzo singularities, known as three-block exceptional
collections. These divide up into three groups, with the special
property that the intersection pairing between elements of  the
same block vanishes. The associated quiver diagram thus always has
a triangular structure. Exceptional three block collections for
the ${\cal B}_8$ singularity have been constructed in
\cite{tworussians}.
The one that is closest to our needs is denoted
as type (8.1) in \cite{tworussians}. Unfortunately the actual collection as given in \cite{tworussians}
turns out not to be exceptional -- probably due to a trivial calculation error. 
Instead we will pick collection (8.2) in \cite{tworussians} and apply Seiberg dualities 
(for a short description, see subsection 3.5)
until we end up with a quiver of type (8.1). The resulting  charge vectors of this collection are:

{\textcolor{gray}{
\ba
\label{coll2}
\shabox{\textcolor{black}{
$\begin{array}{ccc}
{\rm ch}({\FF}_i)\, = \, (1, H\! -\! E_i, 0)\; \  \quad 
\qquad \mbox{\scriptsize ${i =1,\, .  \,   ,4}$}\\[6mm]
{\rm ch}(\FF_i)  \, = \, (1,\! -\! K\! +\! E_i , \,1\, ) \quad 
\qquad \mbox{\scriptsize ${i =5,\, .  \,   ,8}$}\\[4mm]
{\rm ch}({\FF}_9) \, =  \, (1, 2H\! -\! {\mbox{$\sum\limits_{i=1}^4$}} E_i, 0) \qquad \qquad \ \ \ \;
\end{array}$}}\nonumber \\[-4.2cm]
& & \qquad   \shabox{\textcolor{black}{${\rm ch}(\FF_{11}) =\,(6, \! -3K\! 
+ \! \mbox{$2\sum\limits_{i=5}^8 E_i$},  \textstyle{1\over 2})$}}
 \nonumber\\[2mm]
&&   \qquad
\shabox{\textcolor{black}{$
{\rm ch}(\FF_{10}) \, = \,  (3, -K + \mbox{$\sum\limits_{i=5}^8 E_i$}, -\textstyle{ 1\over 2} )$} }
 \nonumber \\
\ea }}
We see that all fractional branes have non-zero
D7 and D5-brane components. The D5 branes are wrapped around the
2-cycles as indicated. From this collection, we wish to obtain the
quiver gauge theory associated with a single D3-brane.
The condition (\ref{nsum})
that all charge vectors must add up to (0,0,1) yields the following multiplicities
\be
n_i = 1\, , \  \ \mbox{\footnotesize ${i =1,\, .  \, ,9}$}
\, , \qquad\qquad  n_{10} =\, 3 \, , \qquad \qquad n_{11}= -3
\ee
So the gauge theory on the D3-brane has gauge group $U(3)^2 \times U(1)^9$.

To obtain the matter content we must determine the dimension of the relevant Ext groups.
Since for each pair of sheaves $\FF_i$ and $\FF_j$ of an exceptional collection, only one
of the Ext groups is non-zero,
one can determine its dimension by computing the corresponding Euler character
\be
\chi(\FF_i,\FF_j) = \sum_k (-)^k {\rm dim \ Ext}(\FF_i,\FF_j)
\ee
which can be computed using the Riemann-Roch formula
\be
\label{euler}
\chi(F_i, F_j) = \int_{\cal B}  {\rm ch}(F_i^*)\, {\rm ch}(F_j) \, {\rm Td}({\cal B}).
\ee
Here ${\rm ch}(F_i)= ({\rm rk} + c_1 + {\rm ch}_2)(F_i)$ denotes the Chern character of $F_i$
and ${\rm Td}({\cal B}) = 1 - {1\over 2} K + H^2$ is the Todd class of the base ${\cal B}$.
For exceptional collections, this formula gives an upper
triangular matrix with all 1's on the diagonal.
Hence we loose no information by anti-symmetrizing:
\be
\label{pair}
\chi_-(\FF_i,\FF_j) = \chi(\FF_i,\FF_j) - \chi(\FF_j,\FF_i) =
{\rm rk}(\FF_i) {\rm deg}(\FF_j) - {\rm rk}(\FF_j) {\rm deg}( \FF_i)\, .
\ee
where ${\rm deg}(\FF_i)$ is the degree of the sheaf, defined as the intersection between the
first Chern class of $\FF_i$ with the canonical bundle of the del Pezzo surface
\be
\label{degree}
{\rm deg}( \FF_i) = - c_1(\FF_i)\cdot K\, .
\ee
Formula (\ref{pair}) counts, with orientation, the number of intersections within ${\cal X}$ 
between the 2-cycle components of one sheaf with the 4-cycle component of the other. Geometrically,
one expects the massless open string states to appear whenever two branes intersect
at a point. In terms of the quiver gauge theory, the matrix $\chi_-$ indeed represents
the adjacency matrix that counts the number of
lines between the nodes. Moreover
we see that it can be computed very simply from the charge vectors of the fractional branes.

We have outlined the procedure
for obtaining the quiver data for a given exceptional collection of fractional branes $\FF_i$. 
The ensuing rules for drawing the quiver diagram are (a) through (e) given in the previous section.
Rules (a) and (b) are  clear, and rules
(d) and (e) represent the special conditions that define an exceptional collection.
Rule (c) is a consequence of the geometric fact that each fractional brane consists
of 0, 2, or 4-cycles only,  and therefore has, on the 6-manifold ${\cal X}$, zero intersection
with a 0-cycle, i.e. with some isolated point $p$. In other words, the intersection pairing
between $\FF_i$ and the sky-scraper
sheaf ${\cal O}_p$ that represents a D3-brane  located at $p$ vanishes.
Using (\ref{nsum}), this implies
\be
\sum_j n_j \, \chi_-(\FF_i,\FF_j) = 0
\ee
for all $i$.  This is rule (c). It in particular ensures that, for the case of
multiple D3-brane probes, each node is free on non-Abelian gauge anomalies.

We can now obtain the full quiver gauge theory for the collection (\ref{coll2}).
Using (\ref{pair}), we obtain the intersection numbers
\ba
\chi(\FF_{11}, \FF_i)  \is  1\, , \nonumber \\[-3mm]
& & \qquad \quad \ \  \  \ \mbox{\footnotesize ${i =1,\, .  \, ,9}$} \nonumber\\[-2mm]
\chi(\FF_i, \FF_{10})  \is \, 1 \,  ,\\[2.5mm]
\chi(\FF_{10}, \FF_{11}) \is 3.\nonumber
\ea
The resulting quiver diagram is given in fig 5. We recognize the characteristic form of
a quiver gauge theory that follows from a three-block exceptional collection.
\medskip
\bigskip

\begin{figure}[htbp]
\begin{center}
\leavevmode\hbox{\epsfxsize=6cm \epsffile{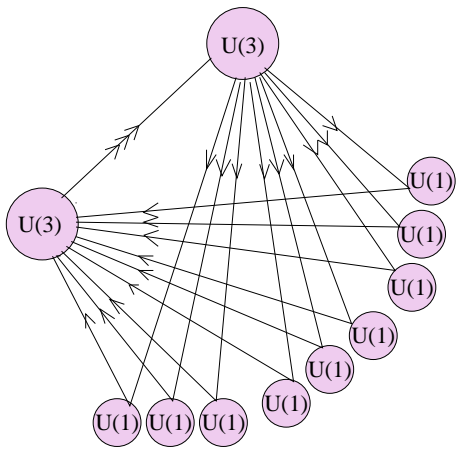}}\\[3mm]
\vspace{2mm}
\caption{ Quiver diagram of the D3-brane gauge theory on a del Pezzo 8 singularity, corresponding
to the exceptional collection of fractional branes given in eqn (\ref{coll2}). This is the
same quiver diagram as that of the D3-brane theory near a ${\bf C}^3/\Delta_{27}$ orbifold singularity.
The connection between the two theories is explained below.}
\end{center}
\end{figure}

\bigskip

\newsubsection{D3-brane on ${\bf C}^3/\Delta_{27}$ orbifold}

\smallskip

As it turns out, the above quiver diagram is identical to that of the D3-brane theory on
the ${\bf C}^3/\Delta_{27}$  orbifold singularity.  Let $X,Y,Z$ denote the three complex coordinates on ${\cal C}^3$. The discrete group $\Delta_{27}$ is the non-abelian subgroup of $SU(3)$  generated by
the three ${\bf Z}^3$ transformations
\ba
\label{gggg}
g_1 \,: \qquad
(X, Y , Z) & {\longrightarrow} & (\, \omega \, X\, , \, \omega^2 \, Y\, , Z\, )
\nonu
g_2 \,: \qquad
(X, Y , Z) & {\longrightarrow} & (\, X, \omega \, Y\, , \omega^2 \, Z\, )
\\[1.5mm]
g_3 \, : \qquad (X,Y,Z)
 & \longrightarrow & \;  (\; Z\; ,\; X\; ,\; Y\; )\nonumber
\ea
with $\omega = e^{\frac{2\pi i}{3}}$. 
The D3-brane theory near this singularity, derived via rules outlined below, 
has the same quiver data as in fig 5.
It is therefore a natural conjecture that del Pezzo 8 singularities can be
viewed as a deformation of this orbifold. This correspondence may be of some use,
since, unlike string theory on a general del Pezzo surface, the worldsheet CFT of strings
on flat space orbifolds is soluble and the D-brane boundary conditions are exactly known
\cite{ALE,digo}.  For completeness, we briefly summarize how the above quiver data
arise from the orbifold construction \cite{albion,Muto}.

Let $\Gamma$ be a general discrete group that acts on ${\bf C}^3$. Consider the D3-brane
and all of its images under $\Gamma$.  Their world-volume theory  is a $U(|\Gamma|)$ gauge
theory with a vector multiplet $V$ and three chiral multiplets $\Phi_i$, that parametrize the
transverse positions of the D3-branes along  ${\bf C}^3$. The orbifold projection amounts to
the requirement that
\ba
\label{vproj}
R_{reg} V R_{reg}^{-1}\is V \nonumber \\[3mm]
(R_3)_{ij} R_{reg} \Phi^j R_{reg}^{-1} \is  \, \Phi^i \label{phiproj}
\ea
where $R_{reg}$ is the regular representation of $\GG$ acting on the
Chan-Paton index, and $R_3$ is the 3-d defining representation.
Since $R_{reg}$ decomposes into irreducible representations as
\be
\label{decomp}
\qquad R_{reg}=\bigoplus_{a=1}^r n_a R^a
\qquad \qquad n_a={\rm dim} R^a.
\ee
the projection (\ref{phiproj}) breaks the  gauge symmetry to $\prod_{a=1}^r U(n_a)$.
Translated into geometric language,
we conclude that a D3-brane near an orbifold singularity splits up into fractional branes $F_a$,
where $a$ labels an irreducible representation $R_a$, and that each fractional brane
occurs with multiplicity $n_a = {\rm dim} R_a$.
The number of chiral fields
$n^3_{ab}$ transforming in the $(n_a, \overline{n}_b)$ bi-fundamental representation, is obtained
by the decomposition
\be
R_3 \otimes R^a = \bigoplus_{b=1}^r n^3_{ab} R^b.
\label{couplings}
\ee

The group $\Delta_{27}$ has 27 elements, that split up in 11 conjugacy classes. It also has
11 representations: nine 1-dimensional representations, and two 3-dimensional ones.
The above orbifold procedure thus produces a quiver  gauge theory with gauge group
$U(3)^2 \times U(1)^9$. Using the formula (\ref{couplings}), a straightforward calculation \cite{Muto} shows
that the bifundamental matter organizes as in the quiver of fig 5.
We are thus motivated to look for a relationship between
the geometry of the orbifold space ${\bf C}^3/\Delta_{27}$ and del Pezzo 8 surfaces.
Consider the following combinations of coordinates
\ba
\label{combi}
x  \is  \, X Y Z  \nonu
y \! \is \! (X^3 \!+ \omega Y^3 \! + \omega^2 Z^3)(X^3 \!+  \omega^2 Y^3\! +  \omega Z^3)\nonu
z  \is X^3
\! + Y^3 \! + Z^3
\\[1mm]
w \! \is\! (X^3\! + \omega Y^3\! + \omega^2 Z^3)^3 \nonumber
\ea
with $\omega = e^{2\pi i \over 3}$. From eqn (\ref{gggg}), it is evident that these expressions
are all invariant under the action of $\Delta_{27}$. Each thus defines a single-valued
coordinate on the orbifold space. If we give $(X,Y,Z)$ weight ${1\over 3}$, then the new invariant combinations in (\ref{combi}) are homogeneous of weight (1,1,2,3). These are the same weights
as of the projective space used in representing ${\cal B}_8$.
With not too much extra work, one can indeed prove that the coordinates $(x,y,z,w)$ defined in
(\ref{combi})  satisfy a homogeneous equation of the form
\be
w^2 + y^3 - 27 w x^3  + w z^3  -3 w yz = 0
\ee
This confirms the identification of ${\bf C}^3/\Delta_{27}$ as a special point in
the moduli space of del Pezzo 8 singularities, and (at least partially) explains the
correspondence of the D3-brane gauge theories.
The orbifold perspective can be useful in case one wants to verify properties of the string
theory using an exact string worldsheet calculation. The general geometric description
of D3-branes is limited to the large volume
regime. On the other hand, as we will see shortly, it has the advantage of being a step closer
to providing a purely geometrical description of the space of gauge invariant
coupling constants. Ideally, of course, one would like to have both descriptions available.

\bigskip

\newsubsection{Seiberg dual}

For a given geometrical singularity, there are in principle many different
exceptional collections of fractional branes. The allowed choices are typically
inequivalent, and in particular lead to different world-volume gauge theories on
the probe D3-brane. There exists a simple transitive set of transformations on the
space of exceptional collections, known as mutations. A useful subclass of
mutations has the physical interpretation of Seiberg duality \cite{Seiberg,freddy}: the
${\cal N}\!=\! 1$ supersymmetric gauge theories corresponding to the original and
mutated set of fractional branes are each others Seiberg dual. For a given singularity,
the question of which of the dual descriptions is most appropriate is determined by the value of
the geometric moduli that determine the gauge theory couplings.

To apply this duality map to a given exceptional collection, one
chooses a particular node $F_i$. One then orders all $F_j$ such that
all branes connected to $F_i$ via incoming lines are placed to the left of $F_i$
and all others are placed to the right. Seiberg duality, applied to the node
$F_i$ then amounts to the following map on the charge vectors situated to
the left of $F_i$
\be
\label{gggo}
{\rm ch}(F_j) \; \longrightarrow \;  {\rm ch}(F_j) - \chi(F_j,F_i) \, {\rm ch}(F_i)
\ee
As a result of this change of basis, the multiplicity of the node $F_i$ needs to be
adjusted, so as to preserve the requirement that all charge vectors must add
up to that of a single D3-brane. The proper adjustment is
\be
\label{klkl}
n_j \; \to  \; n_j -  N_{j}
\ee
where the integer $N_j$ is given by the sum
\be
N_{j} = \sum_{i<j} \chi(F_i,F_j)\, n_j .
\ee
We recognize $N_j$ as the number of flavors at the node  $F_j$. Eqn (\ref{klkl})
thus corresponds to the replacement of $N_c$ with $N_f -N_c$. This supports
the interpretation of the map (\ref{gggo})  as a Seiberg duality.  It is also straightforward
to verify that the change due to (\ref{gggo})  in the number of bi-fundamentals between the
nodes is completely consistent with this physical interpretation.

Geometrically, the transformation (\ref{gggo}) on the basis of charge vectors can be recognized as
the Picard-Lefschetz monodromy around a conifold point. There is a natural interpretation of this.
The quiver theory we have discussed lives at a locus in K\"ahler moduli space where the del Pezzo
surface has shrunk to zero size, but where string perturbation theory is still applicable. There are
other places in K\"ahler moduli space where some cycle has shrunk to zero size and  string
perturbation theory breaks down -- these are generalized
conifold points. We can imagine traversing a loop in moduli space starting at the point where the conformal quiver
theory lives, and going around a conifold point, where the central charge of a given
fractional brane $\FF_i$ vanishes. This will implement the transformation
(\ref{gggo}) on the charge vectors. From the point of view of the worldvolume theory, 
the change in K\"ahler parameters translates into a change in the gauge
coupling of the $U(|n_i|)$ gauge group. As we go around the loop, this gauge coupling is 
pushed through strong coupling, and we have to do a Seiberg duality on the 
$i$th node.\footnote{One should take care to pick a path in moduli space
such that the low energy gauge theory is still applicable and we do not have to worry about 
stringy corrections. It is not
completely clear that this is always possible, but since the effect of monodromy can be d
escribed purely in field theoretic
terms, this seems quite reasonable.}

\newcommand{\ag}{\alpha}
\newcommand{\bg}{\beta}
\newcommand{\cg}{\gamma}

\medskip
\begin{figure}[t]
\begin{center}
\leavevmode\hbox{\epsfxsize=6cm \epsffile{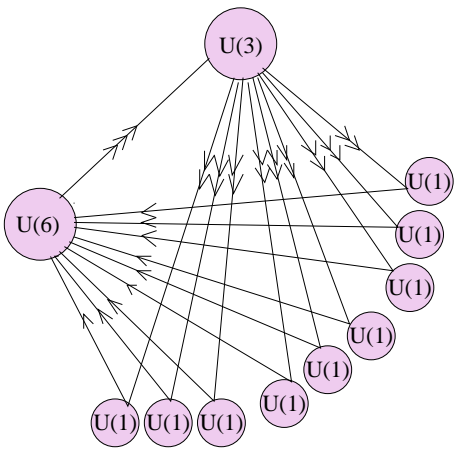}}\\[3mm]
\caption{ \textcolor{black}{Seiberg dual of the D3-brane gauge theory on the $\Delta_{27}$ orbifold, 
or equivalently,
of the exceptional collection of fractional branes (\ref{coll2})-(\ref{coll3}) on the del Pezzo 8 singularity.}}
\end{center}
\end{figure}

Let us specialize to the D3-brane theory on del Pezzo 8.
Besides the $\Delta_{27}$ orbifold quiver theory of fig 5, we now know that it gives rise
to a more general family of ${\cal N} \! = \! 1$ gauge theories obtained via Seiberg dualities. In particular, we can apply Seiberg duality map to the fractional brane $F_{10}$ in eqn (\ref{coll2}).
Via (\ref{ggg}), this map amounts to replacing the neighboring
node $F_{11}$ by a new fractional brane $\tilde{F}_{11}$
with charge vector
\ba
\label{coll3}
{\rm ch}(\tilde{\FF}_{11}) \is
- {\rm ch}(F_{11}) +  3\, {\rm ch}(F_{10})\, =
\, 
(3,  \mbox{$\sum\limits_{i=5}^8 E_i$},  -2)
\ea
As a consequence of this mutation,  the multiplicity of $F_{10}$ changes from
$n_{10}=3$ to $\tilde{n}_{10} = -6$. This is as expected from the Seiberg duality map 
on the field theory: the original node
has 9 flavors and 3 colors, and the new node therefore has $N_f-N_c =6$ colors.
The dual quiver diagram is given in fig 6.
In the next section we will use this Seiberg dual quiver theory as our starting point for
a open string construction of an MSSM-like gauge theory.

\medskip

\newsubsection{Superpotential}

Thus far, we have focused on the topological properties of the D-brane theory. Non-topological
data are harder to control and compute. There is however one more valuable piece of information
that can be extracted with precision from the geometric perspective, namely the holomorphic
superpotential $W$.  For quiver gauge theories, $W$ is a sum of gauge invariant traces
over ordered products of bi-fundamental chiral fields. In principle there is one such term for each
oriented closed loop on the  quiver. In the example of fig 5 or 6, it is known
that $W$ is
a purely cubic function:
\be
W = C_{abc}{\rm Tr}( \phi^a \phi^b \phi^c)
\ee
We can compute the cubic couplings by computing disk three-point amplitudes
in topological string theory. In the large volume limit, the internal part of the vertex operator
for a chiral field is a generator of the Ext group between two fractional branes.
The three-point functions are then proportional to the Yoneda composition of the Ext generators:
\be
{\rm Ext} ^l(i_* F_i,i_* F_j) \times {\rm Ext}^m(i_* F_j,i_* F_k) \times 
{\rm Ext}^{3-l-m}(i_* F_k,i_* F_i)
\to {\rm Ext}^3(i_* F_i,i_* F_i) \equiv {\bf C}.
\ee
This calculation
was done explicitly for del Pezzo singularities ${\cal B}_n$ with $n\leq 6$ in \cite{martijn}.
The superpotential resulting from this calculation is a meromorphic function of the space of
complex structure deformations
of the del Pezzo singularity.

Now let us specialize to the D-brane on ${\cal B}_8$. Let  us first consider
fig 5. Its superpotential $W$ is a cubic expression with $3 \times 9 = 27$ terms,
equal to the number of triangles of the quiver.  Naively this gives 27 independent
Yukawa couplings. However, since we have no direct knowledge of K\"ahler potential
terms, we are free to perform arbitrary field redefinitions, as long as they are
compatible with the structure of the quiver. The group  of allowed field redefinitions is
$GL(1)^9 \times GL(3)$. This reduces the number of independent
parameters in $W$ to $27 - (9 + 9 ) +1 =  10$. (We subtracted the overall scale of $W$.)
The Seiberg dual theory of fig 6 yields to the same number parameters:
there are $3 \times 18 = 54$ terms, but
the group  of allowed field redefinitions is
$GL(2)^9 \times GL(3)$. This again
gives $54 - (9 \times 4 + 9 ) +1 =  10$ parameters.
Eight these 10 parameters can be identified with complex structure deformations of the del
Pezzo 8 surface.



\medskip

\newsubsection{Other geometric moduli}

\smallskip

Besides the superpotential, the D3-brane gauge theory has many other gauge invariant couplings,
which arise as closed string modes of the del Pezzo singularity. These couplings are of vital
importance for making a realistic model.
We will make a few comments about the correspondence below, but the full dictionary has not yet been
established.

The complex structure parameters of a general Calabi-Yau surface are associated to
$(2,1)$ forms and fit into ${\cal N}\! =\! 2$ vector multiplets. For the del Pezzo singularities,
the complex structure moduli that preserve the singularity  correspond to $(2,1)$ forms
with non-compact support.\footnote{There are other complex structure deformations of the Calabi-Yau, which are
localized around the tip of the cone, that make the geometry less singular. They play a role
when one considers worldvolume theories of fractional branes which confine; these
complex structure parameters are then identified which gaugino condensates.}
Their auxiliary fields correspond to turning on various
3-form RR and NS fluxes proportional to the same $(2,1)$ forms and their complex conjugates.
In the 4-dimensional Lagrangian, vector multiplet moduli
appear as spurion fields in the superpotential, and turning on auxiliary fields
gives rise to certain soft supersymmetry breaking terms, namely non-supersymmetric Yukawa couplings 
\cite{albionjohn}.

To every 2-cycle in the Calabi-Yau manifold we can associate a hypermultiplet.
For the $n$-the del Pezzo singularity,
there are $n+1$ 2-cycles. If we denote the two-cycle by $C_I$, then the scalars in the
hypermultiplet are given by the period integrals of two-form potentials. When we put in branes,
we break half of the supersymmetries and we get two ${\cal N}=1$ multiplets for each 2-cycle.
The closed string scalars re-arrange themselves as follows. Consider a D5 brane wrapped
on the 2-cycle. Its 4-d gauge coupling is given by
\be
\label{ggg}
\tau_I 
= \int_{C_I} (C_{(2)}^{RR} - \tau B^{NS})  .
\ee
The other hypermultiplet scalars
\be
\int_{C_I} (J + i C_{(4)})
\ee
become linear multiplets in four dimensions. They contain the Fayet-Iliopoulos terms,
and control the size of the 2-cycles. Thus hypermultiplet moduli appear as spurion fields
in the D-terms. The auxiliary fields of the hypermultiplets
should correspond to some interesting deformations of the background that
to our knowledge have not been studied in any detail yet. Turning them on gives rise to
several interesting soft SUSY-breaking terms, such as gaugino masses and other non-supersymmetric
mass terms.

\bigskip
\bigskip

\newsection{Building the Standard Model on a D3-brane}

We will now look for suitable symmetry breaking process, such that the left-over low energy theory, ideally,
looks like the supersymmetric Standard Model, or as a slightly more modest target, like its minimal
quiver extension: the MQSSM quiver gauge theory introduced in section 2.
The concrete plan is as follows (see fig 7): starting from the D3-brane theory on ${\cal B}_8$,
we choose a suitable configuration of expectation values of bi-fundamental scalar fields. 
We then  translate the symmetry breaking process in the
geometric language of bound state formation of fractional branes.
In this way we manufacture a new basis of fractional
branes, characterized by their collection of charge vectors, such that the corresponding
D3-brane gauge theory reproduces the MQSSM theory. The new basis of fractional
branes will live on a partially resolved del Pezzo 8 singularity.

The minimal quiver theory of a D3 on ${\cal B}_8$,  as given in fig 5,
was the starting point for a string construction of the MSSM proposed in \cite{bjl}.
It was argued  in \cite{bjl} that, by turning on FI-terms, one can
induce the condensation of bi-fundamentals that connect 3 of the
$U(1)$-factors with one of the $U(3)$ nodes, thereby breaking
$U(3)$ to $U(2)$.
Inspection of the full set of  D-term equations, however, shows that one can not
break one of the $U(3)$'s without also breaking the other. This is not what we want,
since we need an unbroken $SU(3)$
color group. For this reason, we will start from the Seiberg dual theory with gauge group
$U(6) \times U(3) \times U(1)^9$. Its quiver diagram is
drawn in fig 6.

\definecolor{darkblue}{rgb}{0,0,.5}
\begin{figure}[t]
\bigskip
\begin{center}
\leavevmode\hbox{\epsfxsize=14cm \epsffile{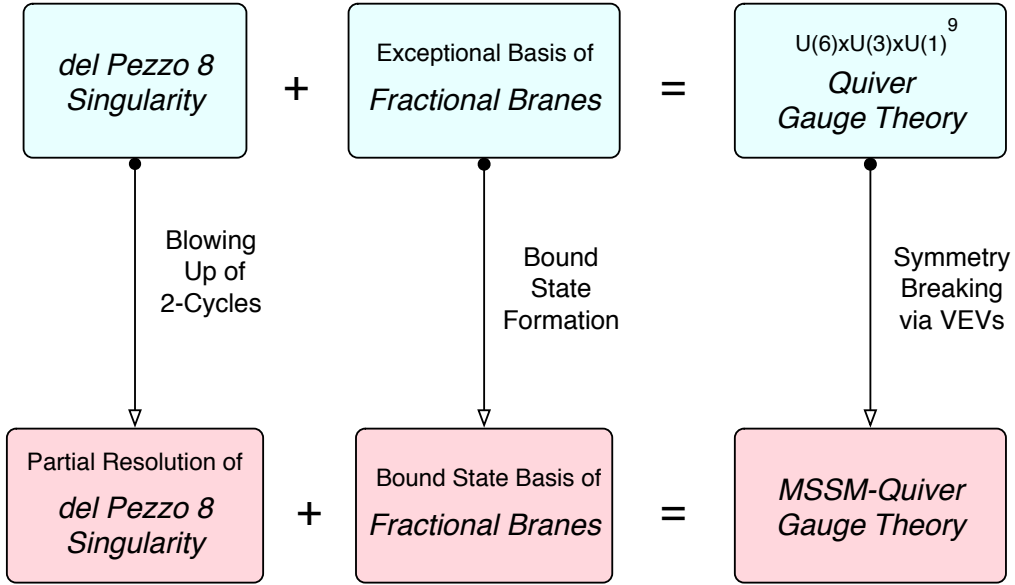}}\\[3mm]
\caption{\textcolor{black}{The ouline of our construction of a geometric dual of the MQSSM.}}
\end{center}
\bigskip
\end{figure}

We will now show that the quiver gauge theory of fig 6 can indeed be reduced
to the MQSSM quiver theory of fig 2. Our main assumption will be that we have
complete freedom to tune all gauge invariant coupling constants: the FI-terms,
the Yukawa couplings, as well as the gauge couplings.

\medskip

\newsubsection{Symmetry breaking to $U(3) \times U(2) \times U(1)^7$}

We denote the three types of bifundamental fields, as
\be
X^p \in (1, \bar 6) \, , \qquad \qquad Z^q \in (6, \bar{3}) \, , \qquad \qquad U^{p,r}\in (3, \bar{1})
\ee
The label  $p$ runs from 1 to 9,  and $q$ runs from 1 to 3, while $r$ runs from 1 to 2.
The general form of the superpotential is
\be
\label{dubya}
W = \sum\limits_{p,q,r} C_{pqr} \, 
X^p Z^q U^{p,r} 
\ee
and the abelian D-term equations are
\ba
\label{abeld}
\qquad \qquad \mbox{$\sum\limits_r$}\, |U^{p,r}|^2 \, -\;  |X^p|^2  \is \zeta_p  \qquad {\mbox{\footnotesize $p=1,..,9$}} \nonumber\\[3mm]
\mbox{$\sum\limits_p$} \, |X^p|^2 - \mbox{$\sum\limits_q$}\, | Z^q|^2\is \zeta_{10}  \\[3mm]
\mbox{$\sum\limits_q$} \, |Z^q|^2 - \mbox{$\sum\limits_{p,r}$} \, |U^{p,r}|^2 \is \zeta_{11}
\nonumber
\ea
Now by assumption,  we allow general deformations of the closed string background,
and we are thus free to arbitrarily tune the couplings in the superpotential, as well as the
FI-terms. For the moment we will assume that the superpotential vanishes, so that we
can ignore the F-flatness conditions. The expectation values of the scalar fields are then
determined via the above equations (\ref{abeld}) in combination with the non-abelian
D-term equations (here $T^a$ and $t^b$ indicate the $SU(6)$ and $SU(3)$ generators)
\ba
\label{dnon}
\mbox{$\sum\limits_p$}\,\, \bar{X}^p\, T^aX^p \is \mbox{$\sum\limits_q$}\,\, \bar Z^q \, T^a Z^q\nonumber\\[3mm]
\mbox{$\sum\limits_q$} \,\,  \bar{Z}^q \; t^b\, Z^q \is \mbox{$\sum\limits_{p,r}$}\, \, \bar U^{p,r}\, t^b U^{p,r}
\ea

After turning on the FI-parameters, the abelian D-term equations dictate that at least some
of the bi-fundamental fields condense. The non-abelian D-flatness equations (\ref{dnon})
 make clear that the condensation must simultaneously occur in all three groups of
 bi-fundamentals. We choose the following special form of the expectation values
 (here we only write the fields with non-zero VEVs):
\ba
\label{row}
X^1 =  {\mbox{\footnotesize $ \left( \begin{array}{c}  { \phi}_1
\\
0
\\  0 \\ 0 \\ 0 \\ 0  \end{array}\right)$}}\, ,\
\qquad \qquad
X^{2}=  {\mbox{\footnotesize $ \left( \begin{array}{c} 0
\\
{\phi}_2
\\  0 \\ 0 \\ 0 \\ 0  \end{array}\right)$}}\, ,
\qquad \qquad X^{3} = {\mbox{\footnotesize $ \left( \begin{array}{c}  0
\\
0
\\  \phi_3  \\ 0 \\ 0 \\ 0  \end{array}\right)$}}\, , \qquad 
\nonumber
\ea
\vspace{-5mm}
\ba
\label{row2}
U^{1,r}\, = \, (\,  \chi^r, \mbox{\small 0, 0}\, )
\ea
\vspace{-5mm}
\ba
\label{row3}
Z^1 = {\mbox{\footnotesize $ \left( \begin{array}{ccc}  {{\psi}_1} & 0 & 0
\\
0 & 0 & 0
\\  0 & 0 & 0 \\ 0 & 0 & 0 \\ 0 & 0 & 0 \\ 0 & 0 & 0  \end{array}\right)$}}\,\qquad \qquad
Z^2 = {\mbox{\footnotesize $ \left( \begin{array}{ccc}  0 & 0 & 0
\\
{{\cal \psi}_2}  & 0 & 0
\\  0 & 0 & 0 \\ 0 & 0 & 0 \\ 0 & 0 & 0 \\ 0 & 0 & 0  \end{array}\right)$}}\,\qquad \qquad
Z^3 = {\mbox{\footnotesize $ \left( \begin{array}{ccc}  0 & 0 & 0
\\
0 & 0 & 0
\\  {{\cal \psi}_3}  & 0 & 0 \\ 0 & 0 & 0 \\ 0 & 0 & 0 \\ 0 & 0 & 0  \end{array}\right)$}}\,
\qquad \nonumber
\ea
This choice breaks the gauge symmetry to
\be
\label{newg}
U(3) \times U(2) \times U(1)^7
\ee
One can always adjust the FI-parameters $\zeta_i$ such that this choice of expectation
values solves the abelian D-term equations. It is not difficult to show that the non-abelian
D-term equations (\ref{dnon}) are also satisfied, for suitable choice of $\phi_n$, $\psi_n$ and $\chi_r$
(see Appendix).

After turning on the superpotential, the F-flatness equations will impose additional restrictions.
For a general choice of $W$, these may not be solved by the above configuration
of vacuum expectation values. In this case, the theory may need to choose another symmetry
breaking pattern or supersymmetry may be broken. We will assume, however,
that we have sufficient control over all parameters to ensure that
the above expectation values are compatible with F-flatness.  Specifically,
we assume that all Yukawa couplings that connect two fields with a non-zero VEV
can be tuned to zero. This amounts to the six conditions (for notation see eqn (\ref{dubya}))
\ba
\label{special}\qquad
C_{q,q,r} = 0
\qquad \quad
 \mbox{\scriptsize{$q=1,2,3$; \  $r=1,2$}}.
\ea
The remaining non-zero Yukawa couplings then typically result in mass-terms
for the matter fields, proportional to their coupling to the vacuum condensates.
We would like to determine the typical matter content that survives in the low-energy
theory. We could of course proceed to study this question from
the gauge theory point of view. Instead, however, let us first return to the geometric description
in terms of fractional branes, since this provides a useful dual perspective.

\medskip

\newsubsection{Geometric derivation of the low energy theory}

\smallskip

The unbroken gauge theory corresponds to the exceptional collection of fractional branes
with charge vectors as given in (\ref{coll2}), with $F_{11}$ replaced by $\tilde{F}_{11}$
in (\ref{coll3}). In order to trigger the symmetry breaking we switched on
certain FI-terms; by turning on large VEVs and integrating out very massive modes, we reduce
to a simpler quiver theory. From the geometric perspective, turning on FI-terms corresponds
to turning on certain blow-up modes which partially resolve the singularity. The Higgsed down
quiver theory is the worldvolume theory for a D3-brane probing this simpler, partially resolved
singularity.

A basis of fractional branes for the simpler singularity may be obtained from
the fractional branes of the original singularity. The intuitive picture is that turning on VEVs
in the quiver theory corresponds to bound state formation of fractional branes. Of course,
we have been using this idea all along, because our quiver theory is just a way of describing our
probe D3 brane as a bound state of fractional branes.

To describe this condensation process in terms of sheaves on collapsing 
cycles\footnote{Or complexes of such sheaves.}
can be somewhat complicated. However it can be very simply described at the level of charge vectors.
When two fractional branes $\FF_1$ and $\FF_2$ bind into $\FF_B$, the corresponding  nodes in the quiver
diagram collapse to one.
The charge vector of the bound state associated to the new node is the sum of the two constituents
\ba
{\rm ch}(\FF_B) \is {\rm ch}(\FF_1) +  {\rm ch}(\FF_2)\, ,
\ea
So it is relatively straightforward to obtain the charge vectors associated to each node in the
new quiver diagram, and hence this is a simple method to determine the net field content after
condensation.

The pattern of bound state formation in our model follows by inspection of the set
of expectation values (\ref{row2}). The rule  we will follow is that all fractional branes
(= gauge group factors)
that are connected by matter fields with a non-zero expectation value are part of the same bound state.
Applying this rule, we arrive at the following charge vector of the bound state:
\ba
\label{bds}
{\rm ch}(F_{0}) \is 3 \, {\rm ch}(F_{10})
-\mbox{$\sum\limits_{i=1,2,3}$}{\rm ch}(F_i)  - {\rm ch}(\tilde{F}_{11})
\nonumber
\ea
\vspace{-5mm}
which gives
\ba
\label{finally}
{\rm ch}(F_{0})
\is (3, -\! 2 K \! +\! \mbox{$\sum\limits_{i=5}^8 E_i$}\! - E_4,  {\textstyle{1\over 2}} )
\ea
As a result, the new basis of fractional branes is $(F_0,F_4,\ldots, F_9, F_{10}, \tilde{F}_{11})$.
The respective multiplicities are
$(1,1, \ldots, 1,-3,2)$, in accordance with (\ref{newg}).

The number of oriented lines that connect to the new node are obtained by
adding or subtracting, depending on the relative orientation, all the lines
that connect to the original two nodes. These reduction rules  for eliminating nodes
and lines from the quiver diagram properly reflect the lifting of gauge fields
and bi-fundamental matter from the low energy theory. To determine
the matter content, we compute the intersection pairings according to the formula (\ref{pair}).
The respective ranks of the fractional branes are $(3, 1, \ldots, 1, 3, 3)$ and the respective
degrees are $(5,2, \ldots, 2,5,4)$.
We thus obtain the following intersection matrix
\ba
\label{chiminus}
\chi_-(F_i,F_j) \is
\mbox{\footnotesize
$\left(\! \begin{array}{ccccccccc}
\ 0 & \ 1 & \ 1 & 1 & \ 1 & \ 1 & 1 & 0 &  -3 \\
-1 & \ 0 & \ 0 & 0 & \ 0 & \ 0 & 0 & -1 &  -2\\
-1 & \ 0 & \ 0 & 0 & \ 0 & \ 0 & 0 & -1 &  -2 \\
-1 & \ 0 & \ 0 & 0 & \ 0 & \ 0 & 0 & -1 &  -2 \\
-1 & \ 0 & \ 0 & 0 & \ 0 & \ 0 & 0 & -1 &  -2\\
-1 & \ 0 & \ 0 & 0 & \ 0 & \ 0 & 0 & -1 &  -2 \\
-1 & \ 0 & \ 0 & 0 & \ 0 & \ 0 & 0 & -1 &  -2 \\
\ 0 & \ 1 & \ 1 & 1 & \ 1 & \ 1 & 1 &\ 0 &  -3\\
\ 3 & \ 2 & \ 2 & 2 & \ 2 & \ 2 & \ 2 & \ 3 & \ 0
\end{array}\!
\right) $}
\ea
The resulting quiver diagram is depicted in~fig~8. We organized the bifundamental matter
content into three generations of quarks and leptons. The quiver looks very similar to
the minimal quiver extension  of the MSSM as drawn in fig 2. The only difference is that
two of the $U(1)$ nodes are replaced by $U(1)^3$. We will discuss possible ways of eliminating
these extra $U(1)$'s
in the subsection 4.4.
%

In eqn (\ref{chiminus}),  $\chi_-(\FF_i,\FF_j)$  denotes the anti-symmetric part of the
the Euler character: it
counts the number of bifundamentals from $F_i$ to $F_j$ minus those from $F_j$ to $F_i$.
However, the new basis of fractional branes
is no longer an exceptional collection, and it would therefore be possible that
lines of both orientations appear between two nodes.
To obtain more information, let us compute the full formula (\ref{euler})
for the  Euler character.
\footnote{{Evaluation of the integral (\ref{euler}) gives:
$$
\chi(F_i,F_j) =
{\rm rk}(F_i) {\rm rk}(F_j )\! + \! {\rm rk}(F_i) {\rm ch}_2(F_j) + {\rm ch}_2(F_i)
 {\rm rk}(F_j) \! - c_1(F_i)\cdot c_1(F_J) + {1\over 2} \chi_-(F_i,F_j)$$}}
We obtain
\be
\chi(F_i,F_j) =
\mbox{\footnotesize
$\left(\! \begin{array}{ccccccccc}
\ 1 &  0 &  0 & 0 &  0 &  0 & 0 & 0 &  \!\!-2 \\
-1 & 1 &  0 & 0 &  0 &  0 & 0 & 0 &  0\\
-1 & 0 &  1 & 0 &  0 &  0 & 0 & 0 &  0 \\
-1 &  0 &  0 & 1 &  0 &  0 & 0 & 0 &  0 \\
-1 &  0 &  0 & 0 &  1 &  0 & 0 & 0 &  0\\
-1 &  0 &  0 & 0 &  0 &  1 & 0 & 0 &  0 \\
-1 & 0 &  0 & 0 &  0 &  0 & 1 & 0 &  0 \\
\ 0 &  1 &  1 & 1 &  1 &  1 & 1 & 1 &  0\\
\ 1 &  2 &  2 & 2 &  2 &  2 & 2 &  3 &  1
\end{array}\!
\right) $}
\ee
Let us make
some comments on this result. First, the fact that all diagonal elements are equal to 1 is good news:
the 1 represents the gauge multiplet of the corresponding node, and thus indicates the absence
of any adjoint matter.  We further see that, although this set of fractional branes cannot correspond
to an exceptional collection, the
abundance of 0's in off-diagonal entries shows that it comes very close.
This is desirable, since it gives a strong indication that there are no extra fields beyond the ones
exhibited in the quiver diagram of fig 8.

\newcommand{\ccdot}{{\textcolor{black}{\mbox{\scriptsize $*$}}}}


\bigskip

\begin{figure}[t]
\begin{center}
\leavevmode\hbox{\epsfxsize=9cm \epsffile{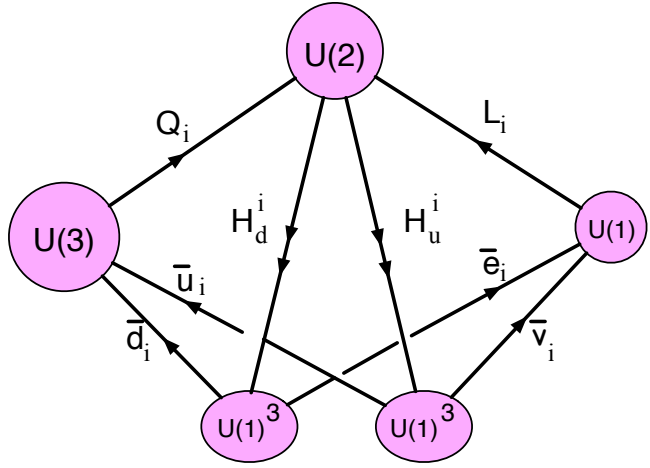}}\\[3mm]
\caption{\textcolor{black}{The quiver diagram of the collection of fractional branes given in eqns (\ref{coll2}) and (\ref{finally}). It has the same matter content and gauge symmetry as the MQSSM
given in fig 2, except that each of the three generations couples to a different $U(1)$ factor.}}
\end{center}
\end{figure}

\newsubsection{A field theory check}

As an independent check on the geometric calculation, we can try to obtain the spectrum and interactions
after symmetry breaking from the field theory perspective. Based on the form (\ref{row2})
of the condensates, as well as the structure of the quiver in fig 8, we propose that the MSSM fields
are obtained from the original quiver fields as follows 
\ba
\label{nrow}
X^i =  {\mbox{\footnotesize $ \left( \begin{array}{c} \; \ccdot \;
\\
\ccdot
\\  \ccdot  \\ \ccdot \\  \ccdot \\ \ccdot  \end{array}\right)$}}\, ,\
\qquad \qquad \ \ \
X^{i+3}=  {\mbox{\footnotesize $ \left( \begin{array}{c}  \bar{\nu}_i 
\\
\ccdot
\\  \ccdot \\ \bar{u}_i \\ \bar{u}_i \\ \bar{u}_i   \end{array}\right)$}}\, ,
\qquad \qquad \ \  \ X^{i+6} = {\mbox{\footnotesize $ \left( \begin{array}{c}  \bar{e} 
\\
\ccdot
\\  \ccdot
\\ \bar{d}_i \\ \bar{d}_i \\ \bar{d}_i  \end{array}\right)$}}\, ,
\nonumber
\ea
\vspace{-6mm}
\ba
\label{nrow2}
\quad U^{i,r} \! =  (\; \ccdot\, , \; \ccdot\, ,\; \ccdot\; )\qquad \qquad
U^{i+3,r} \! = 
(\, \ccdot\,,\mbox{\footnotesize $H^{i,r}_u, H^{i,r}_u$}\, ) 
\qquad \qquad
U^{i+6,r} \! =  
(\, \ccdot\,,\mbox{\footnotesize $H^{i,r}_d, H^{i,r}_d$}\, ) 
\nonumber
\ea
\vspace{-6mm}
\ba
 Z^1 = {\mbox{\footnotesize $ \left( \begin{array}{ccc}  \ccdot & {L}_1  & {L}_1 
\\
\ccdot  & \ccdot & \ccdot 
\\  \ccdot & \ccdot & \ccdot \\ \ccdot &  { Q}_1&  {Q}_1 \\ \ccdot &
{Q}_1 &  {Q}_1  \\ \ccdot & {Q}_1 &  {Q}_1 \end{array}\right)$}}\,
\qquad
\quad
 Z^2 = {\mbox{\footnotesize $ \left( \begin{array}{ccc}
\ccdot  & \ccdot & \ccdot \\ \ccdot & {L}_2  & {L}_2
\\  \ccdot & \ccdot & \ccdot \\ \ccdot &  { Q}_2 &  {Q}_2\\ \ccdot &
{Q}_2 &  {Q}_2  \\ \ccdot & {Q}_2 &  {Q}_2 \end{array}\right)$}}\,
\qquad \quad
 Z^3 = {\mbox{\footnotesize $ \left( \begin{array}{ccc} \ccdot  & \ccdot & \ccdot 
\\  \ccdot & \ccdot & \ccdot \\  \ccdot & {L}_3  & {L}_3 \\
\ccdot &  { Q}_3 &  {Q}_3\\ \ccdot &
{Q}_3 &  {Q}_3  \\ \ccdot & {Q}_3 &  {Q}_3 \end{array}\right)$}}\,  \nonumber
\ea
with $i=1,2,3$.
All entries not indicated with a $*$ have, due to the constraints (\ref{special}) on the
superpotential, no direct Yukawa coupling to the fields with non-zero expectation value.
These components are thus expected to
remain massless after the symmetry breaking. Conversely, all fields marked by the $*$'s
are expected to acquire a mass, either via Yukawa coupling to a field with non-zero VEV or
by being eaten via the Higgs mechanism.

While it would be useful to analyze the symmetry breaking process is more detail, we like to
emphasize that we prefer to view the original high energy quiver theory as only an
intermediate step towards a direct string construction of the unbroken  MSSM-like theory.

\medskip

\newsubsection{Decoupling of $U(1)$ symmetries}

\smallskip

The quiver diagram of fig 8 displays a total of nine $U(1)$ factors. Most of these, however,
are automatically or easily decoupled from the low energy theory.  First, since all fields
are neutral under the overall $U(1)$ symmetry $Q_{tot} = \sum_i Y_i$, this overall factor
decouples. Secondly, it can easily be shown that there are two $U(1)$ factors that have
mixed anomalies with the non-abelian gauge symmetries. In the full string theory realization
of the quiver gauge theory, these anomalies are cancelled via a generalized Green-Schwarz mechanism,
which in addition renders the corresponding $U(1)$ vector bosons massive. In the decoupled low energy
theory, these $U(1)$'s thus survive as anomalous global symmetries. For completeness, let us
briefly outline the relevant stringy cancellation mechanism \cite{mass}.

The world-volume action of fractional Dp-branes includes a CS-coupling to the RR-potentials of the form
\be
\label{rrc}
 \int\! C^{(p-1)}\! \wedge {\rm Tr} F\; , \qquad {\rm and}
\qquad  \int\! C^{(p-3)}\! \wedge {\rm Tr} ( F \wedge F)\; .
\ee
Upon integrating out the $C$-fields, these interaction terms give rise to additional anomalous
contributions to the D3-brane gauge theory action that cancel the quantum mechanical anomalies,
due to the presence of chiral fermions. The compensating contributions arise from the coupling,
via the C-field propagator, between the first and second type of interaction terms in eqn (\ref{rrc}).
The coupling between two interaction terms of the first type in (\ref{rrc}) gives rise to the
Stueckelberg type mass-terms for the vector bosons.
In the decoupled theory, only the gauge bosons of the anomalous $U(1)$ factors acquire a mass.
An intuitive explanation\footnote{It would be worthwhile to work out the 
following argument in more detail.}  for this is that the anomalous $U(1)$
factors are in one-to-one correspondence with fractional branes that wrap cycles that, via
the intersection pairing, are dual to {\it compact} cycles within the non-compact Calabi-Yau
geometry. There are two such cycles: the 4-cycle that wraps the del Pezzo 8 surface, and
its dual degree one 2-cycle. By contrast,
all degree zero 2-cycles within the del Pezzo surface are dual
to non-compact 4-cycles. It is therefore natural that the associated closed string modes,
which are the would-be longitudinal components of the non-anomalous $U(1)$'s, have non-normalizable
kinetic terms. The vector bosons of the anomaly free $U(1)$'s thus survive as massless
low energy degrees of freedom.

This leaves us, of the original nine, with a total of six $U(1)$ gauge symmetries. Their generators
can be recognized as the hypercharge $Y$ and $B- L$, as identified for the MQSSM in section 2,
and four additional charges, given by the difference of two generators within each
$U(1)^3$ node in fig 8. There are several ways in which the extra $U(1)$'s can be decoupled
from the low energy physics. For instance, it is natural to assume that the superpartners of
the right-handed neutrinos acquire a non-zero expectation value. This VEV breaks three 
of the $U(1)$ symmetries. Turning on generic Higgs expectation values will then break all 
the remaining U(1)'s, {\it except} for the electro-magnetic gauge group generated by
$Q_{{}_{EM}} = Y + T_3$. In other words: simply by assuming non-trivial sneutrino and Higgs VEVs,
our model automatically leads to the correct electro-magnetic charge assignments of quark and leptons.

For the purpose of reproducing the MQSSM of fig 2 from our quiver in fig 8, we are free to
adopt a direct approach and decouple the extra $U(1)$ vector bosons from all observable matter, 
simply by tuning the corresponding $U(1)$ coupling 
constants and make them sufficiently small.\footnote{As seen from eqn (\ref{ggg}), 
tuning several gauge couplings to be very small will
necessarily involve a limit in which the string coupling $g_s$ becomes very small, while
simultaneously tuning the periods of the $B$-field to a critical value. A similar limit is used
in the definition of Higgsed little string theories \cite{kutasovea}.}
The extra $U(1)$'s then become global symmetries, that forbid certain undesirable generation mixing
couplings.  
Upon taking this decoupling limit, we obtain precisely the MQSSM discussed in section 2.

\medskip

\newsubsection{Geometric dual of the MQSSM}

\smallskip

With some hindsight, we can now give a more intrinsic geometric characterization
of the reduced geometry and collection of fractional branes described by the MQSSM,
from which any reference to
the symmetry breaking process and bound state formation has been erased.
Notice that  the new bound state
basis of fractional branes, as given in eqns (\ref{coll2}), (\ref{coll3}) and (\ref{finally}), 
does not contain the 2-cycles $\alpha_1$ and $\alpha_2$
(see eqn (\ref{ddd})). This indicates
 that these 2-cycles have been removed from the singularity
 by blowing them up.

\begin{figure}[htbp]
\medskip
\begin{center}
\leavevmode\hbox{\epsfxsize=5.5cm \epsffile{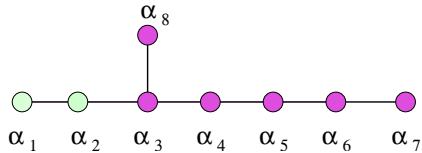}}\\[2mm]
\caption{ \textcolor{black}{The symmetry breaking towards the MQSSM
of fig 8
is geometrically dual to blowing up the 2-cycles $\alpha_1$ and $\alpha_2$,
corresponding to the first two roots of $E_8$, of the del Pezzo 8 surface. In order to perform
the blow-up, the complex structure is tuned so that the del Pezzo develops an
$A_2$ singularity. }}
\end{center}
\end{figure}
Being cycles with self-intersection $-2$, the $\alpha_i$ in general
do not appear as effective curves: they can be blown up only at special points
in the complex structure moduli space, at which the del Pezzo 8 surface develops
a suitable A-D-E type singularity. In our case, we first need to tune the complex
structure to obtain an $A_2$ singularity. This tuning
is presumably the geometric equivalent of the requirement (\ref{special}) on the
superpotential $W$. At this special locus in complex structure moduli space,
the Calabi-Yau singularity
becomes degenerate and the quiver moduli space develops new branches. 
Then, the pair of 2-cycles
$\alpha_1$ and $\alpha_2$ may be
be blown up and  removed from the
singularity. The remaining CY singularity is our proposed geometric dual to the 
MQSSM.\footnote{Since the Standard Model is not a strongly coupled large $N$ gauge theory,
the dual classical geometric description has somewhat limited validity, since 
string corrections
are bound to be important. As a purely theoretical exercise, one could imagine taking a large
$N$ limit of the MQSSM, by considering a large number of D3-branes on the partially resolved
del Pezzo 8 singularity. After accounting for the backreaction and taking a decoupling limit,
this leads to an AdS/CFT type dual geometry, which for large 't Hooft coupling is arbitrarily
weakly curved. We expect this dual geometry to be directly related to our del Pezzo 8 surface
with an $A_2$ singularity, presumably via a geometric transition analogous to \cite{transition}.}

\bigskip
\bigskip

\newsection{Discussion}

We have identified a Calabi-Yau singularity on which the D3-brane world-volume theory
reproduces the MQSSM theory of fig 2. In terms of the quarks, leptons, and 
gauge bosons, it has the exact same matter content as the MSSM. There are however
a number of additional Higgs fields, as well as a possibly a number of (arbitrarily 
weakly coupled) extra $U(1)$ gauge factors. 
In general, the appearence of extra Higgses and $U(1)$'s is characteristic of 
many string theoretic models,
as well as to many other proposals for extending the Standard Model.  In all cases,
it is important to have the freedom to tune masses and couplings, to ensure 
compatibility with current observational bounds. Whether the
present model can deal with these challenges needs further study, but the ability
to (fine-)tune couplings is one of the main strengths of our set-up.

A crucial part of supersymmetric model building is the understanding of how supersymmetry gets
broken. In general, supersymmetry breaking is parameterized by means of soft terms, and
each mechanism generates its own characteristic pattern. In our set-up, a certain class of
soft SUSY breaking terms can be geometrically understood as the effect of turning on IASD
three form flux \cite{soft}\cite{albionjohn}. 
As mentioned earlier in section 3.6, these fluxes appear as auxiliary fields of complex 
structure moduli in the superpotential and thus turning them on gives rise to non-supersymmetric 
Yukawa couplings. Non-SUSY mass terms can be generated by auxilary fields of closed string
hypermultiplets. Unfortunately, their geometric meaning is not well understood at present,
and since they appear in D-terms, their couplings to world-volume fields are much harder
to compute. Nonetheless, it is evidently worthwhile to develop a better control and understanding
of supersymmetry breaking in our set-up. In this respect, an interesting recent development
is the realization that adding an extra vanishing del Pezzo with fractional branes to our 
set-up introduces a hidden sector in which supersymmetry may be dynamically broken \cite{threegroups}.


Finally, let us address some possible criticisms of our bottom-up approach to string phenomenology.
Right from the start, one could ask why it would even be useful to try to
construct the Standard Model via a decoupling limit  on one or more D3-branes.
Since the decoupled theory has continuous parameters, the approach does not
seem to be restrictive enough to lead to a phenomenologically predictive framework
-- at least not much beyond that of ordinary quantum field theory.
Our point of view, however, is that the open string approach is a very reasonable
first step towards the larger goal of string phenomenology. Eventually, QFT breaks
down at the Planck scale, and string theory is our best
chance of finding a fully consistent UV completion. To find out which closed string
theory is the right one, it is useful to know how our present knowledge
of the Standard Model translates into the geometric language of string theory. This is what we
have tried to investigate. An inevitable
hurdle in this quest is that the geometric language has somewhat limited validity in the regime
of interest. As we hope to have shown, however, useful lessons can still be learned by trying
to match the two perspectives.

Our construction of the MSSM-quiver gauge theory of fig 8 depends on
several seemingly arbitrary choices. One can easily imagine arriving at an equivalent
gauge theory via an alternate route, starting from a different geometry and
via a different symmetry breaking process. It seems reasonable, however, that if the final
D3-brane gauge theory is the same, the final geometric singularity must also be the same.
We expect that, in this respect, our bottom-up approach is robust.

What does our model add to the many other D-brane constructions of Standard Model
like field theories \cite{bottomup,urangareview,SMb,bcls}? A key distinction between our model and
almost all other existing proposals, is that in our case all D-branes are localized near
a very small neighborhood of the compactification manifold. 
As a result,  the open string
dynamics can be controllably
disentangled from the dynamics of the closed string moduli. In this way we can
cleanly separate the question of closed string moduli stabilization from that of building
a realistic low energy field theory. We view that as an important advantage of the
bottom-up perspective.

A final important possible objection to our set-up is that it seems to ignore the lesson
 of the unification of coupling constants.  Although, by choosing to work with a single
D3-brane, our specific construction does achieve some form of geometric unification,
we see no obvious reason why, in our model, the couplings would need to
converge at some high energy scale. Gauge unification is indeed somewhat
at odds with the bottom-up philosophy, but we prefer to see the two viewpoints
as complementary rather than incompatible.

\bigskip

\bigskip

\bigskip

\bigskip

\noindent
{\large \bf Acknowledgements}

It is a pleasure to thank Bobby Acharya, David Berenstein, Oliver DeWolfe,
Simeon Hellerman, Luis Ibanez, Paul Langacker, Shamit Kachru, Igor Klebanov, 
Dima Malyshev, David Morrison, Radu Roiban,
Raul Rabadan, Leonardo Rastelli, Nati Seiberg,  Gary Shiu and
Johannes Walcher for helpful discussion and comments. This work
supported by the National Science Foundation under grants No.~0243680. Any opinions,
findings, and conclusions or recommendations expressed in  this material are
those of the authors and do not necessarily reflect the views of the National Science
Foundation.

\bigskip

\appendix

\renewcommand{\newsection}[1]{
\addtocounter{section}{1} \setcounter{equation}{0}
\setcounter{subsection}{0} \addcontentsline{toc}{section}{\protect
\numberline{\Alph{section}}{{ \rm #1}}} \vglue .6cm \pagebreak[3]
\noindent{\large \bf  \thesection. #1}\nopagebreak[4]\par\vskip .3cm}
\renewcommand{\newsubsection}[1]{
\addtocounter{subsection}{1}
\addcontentsline{toc}{subsection}{\protect
\numberline{\Alph{section}.\arabic{subsection}}{#1}} \vglue .4cm
\pagebreak[3] \noindent{\it \thesubsection.
#1}\nopagebreak[4]\par\vskip .3cm}

\newsection{Appendix: Verification of D-term equations}

As promised, let us verify that the expectation values (\ref{row})-(\ref{row3}) satisfy all necessary
D-term equations. We assume that the abelian D-term equations are satisfied by suitable adjustment
of the FI-parameters. The $SU(6)$ D-flatness condition requires that
\be
\sum_{p} \, ({X}^{p})^\dagger \otimes X^p
= {\mbox{\footnotesize $ \left( \begin{array}{cccccc}  |{{\cal \phi}_1}|^2
& 0 & 0 & 0 & 0 & 0\\
0 & |{ \phi_2}|^2 & 0 & 0 & 0 & 0
\\  0 & 0 & |{\phi_3}|^2  & 0 & 0 & 0\\ 0 & 0 & 0 & 0 & 0 & 0 \\ 0 & 0 & 0 & 0 & 0 & 0 \\ 0 & 0 & 0  & 0 & 0 & 0 \end{array}\right)$}}
\ee
equals
\be
\sum_{q,i} \, (Z_i^{q})^{\dagger} \otimes Z^q_i =
{\mbox{\footnotesize $ \left( \begin{array}{cccccc}  |{{\cal \psi}_1}|^2
& 0 & 0 & 0 & 0 & 0\\
0 & |{ \psi_2}|^2 & 0 & 0 & 0 & 0
\\  0 & 0 & |{ \psi_3}|^2  & 0 & 0 & 0\\ 0 & 0 & 0 & 0 & 0 & 0 \\ 0 & 0 & 0 & 0 & 0 & 0 \\ 0 & 0 & 0  & 0 & 0 & 0 \end{array}\right)$}}
\ee
where $i$ denotes the $U(3)$ index of $Z^q$. So we find that this equality is satisfied
provided we choose $\phi_n$ equal to $\psi_n$. Similarly, the $SU(3)$ equations require that
\be
\sum_{q,I} \, (Z_I^{q})^{\dagger} \otimes Z^q_I =
{\mbox{\footnotesize $ \left( \begin{array}{ccc} {\mbox{ $\sum\limits_{q=1}^3$}}
|{\cal \psi}_q|^2  & 0 & 0
\\
 0 & 0 & 0
\\   0 & 0 & 0 \end{array}\right)$}}
 \ee
where $I$ denotes the $U(6)$ index of $Z^q$, equals
 \be
\sum_{p,r} \, (U^{p,r})^{\dagger}  \otimes U^{p,r} =
{\mbox{\footnotesize $ \left( \begin{array}{ccc} {\mbox{$\sum \limits_{r=1}^2$}}
|{ {\chi}_r}|^2
& 0 & 0
\\
 0 & 0 & 0
\\   0 & 0 & 0 \end{array}\right)$}}
\ee
This equality fixes the value of $\sum |\chi_r|^2$. So we conclude that (\ref{row})-(\ref{row3}) indeed represents a valid vacuum, provided the superpotential $W$ is adjusted
appropriately.

\bigskip
\bigskip

\renewcommand{\Large}{\large}

\end{document}